\documentclass{article}

\usepackage{PRIMEarxiv}

\usepackage{abstract}
\usepackage[utf8]{inputenc} % allow utf-8 input
\usepackage[T1]{fontenc}   % use 8-bit T1 fonts
\usepackage{hyperref}       % hyperlinks
\usepackage{url}            % simple URL typesetting
\usepackage{booktabs}       % professional-quality tables
\usepackage{amsfonts}       % blackboard math symbols
\usepackage{amsmath}        % for gather and other math environments
\usepackage{amsthm}         % theorem environments
\usepackage{nicefrac}       % compact symbols for 1/2, etc.
\usepackage{microtype}      % microtypography
\usepackage{lipsum}
\usepackage{braket}
\usepackage{fancyhdr}       % header
\usepackage{graphicx}       % graphics
\graphicspath{{media/}}     % organize your images and other figures under media/ folder

\newtheorem{proposition}{Proposition}[section]

\title{Quantum-inspired Techniques in Tensor Networks for Industrial Contexts
 }

\author{
  Alejandro Mata Ali \\
  Instituto Tecnológico de Castilla y León, Burgos, Spain\\
  \texttt{alejandro.mata.ali@gmail.com} \\
   \And
  Iñigo Perez Delgado \\
  i3B Ibermatica, Parque Tecnológico de Bizkaia \\
  Ibaizabal Bidea, Edif. 501-A \\
  48160 Derio, Spain\\
  \texttt{iperezde@ayesa.com} \\
   \And
  Aitor Moreno Fdez. de Leceta \\
  Quantum Technologies and Systems Unit,\\
  LKS Next, MONDRAGON Corporation, Goiru 7,\\
  20500 Arrasate-Mondragón, Gipuzkoa, Spain\\
  \texttt{aitormoreno@lksnext.com} \\
}

\begin{document}
\maketitle

\begin{abstract}
In this paper we present a study of the applicability and feasibility of quantum-inspired algorithms and techniques in tensor networks for industrial environments and contexts, with a compilation of the available literature and an analysis of the use cases that may be affected by such methods. In addition, we explore the limitations of such techniques in order to determine their potential scalability.
\end{abstract}

\tableofcontents
\section{Introduction}
Quantum computing is a novel field that is receiving considerable interest due to its particular way of performing operations and processing information, being able to tackle highly complex computational problems. In the information and digitalization era in which we find ourselves, this technology has a remarkable applicability for industrial cases, for example in the optimization of delivery routes \cite{TSP_Quantum1,TSP_Quantum2}, packet storage \cite{Bin_Quantum} or quantum machine learning \cite{QML} applied to various contexts \cite{Vision,Quanvolutional,Class1,Class2}.

However, despite the great progress in the development of digital quantum (or `logic gate') computers, substantial challenges remain, such as the fragility of quantum states, errors in quantum operations, or the scalability of the number of qubits.

In this context, there is a need to explore alternative approaches that can take advantage of the computational properties of quantum systems, but do not have to run on digital quantum computers. One such approach may be quantum annealing \cite{Annealing,Dwave}, which runs on specialized quantum devices dedicated to combinatorial optimization of particular cost functions. Another one is digital annealing \cite{Digital,Digital2,Digital3}, which draws on properties of quantum annealing to simulate it classically.

On the other hand, one of the most promising technologies, although less known within industry, is tensor networks (TN) \cite{TN}. Tensor networks are a class of quantum-inspired algorithms and techniques based on mimicking the tensor operations performed by a quantum computer, but executing them on classical computers. By using tensor properties, the execution of such operations can be optimized, especially in cases where the entire quantum state vector is not required, but only properties of it.

Another relevant feature of tensor networks is that they allow the efficient representation of certain families of quantum states, by means of representations such as the matrix product state (MPS)\cite{MPS1,MPS2}, also called tensor train (TT), or projected entangled pair states (PEPS) \cite{MPS2}. In this way, with a reduced amount of memory we can perform quantum calculations and obtain properties of complex systems. These same representations have also been relevant to machine learning, since they can compress models by reducing the memory required without necessarily losing a notable amount of precision \cite{Compress1,Compress2,ML_TN}.

Given all these capabilities, tensor networks are relevant to industrial contexts when the chosen tensorization exposes low-rank, local, hierarchical, sparse, or topology-aligned structure. When the required ranks or contraction widths grow rapidly, the same representations may become impractical. Despite being a relatively mature field at the academic level (initiated in 1971 by Penrose \cite{Penrose}), its study applied to industry is more recent and often associated with quantum computing. At the same time, recent tensor-network simulations have also become part of the classical benchmark used to assess quantum-hardware claims \cite{Q_Adv1,Q_Adv2,QAdvPerspective}, which has increased interest in their practical deployment.

In this paper we survey the main uses of tensor networks for industrial use cases, compiling state-of-the-art articles and exposing the strengths and weaknesses of each technique. In addition, we classify the algorithms according to their specific use case. The field of tensor networks has an extensive and complex literature, so the aim of this article is to serve as an introductory guide to it.

\subsection{Tensor networks in 2026: scope and current role}
As of 2026, tensor networks occupy a broader role than the classical simulation of isolated quantum systems. They are used as compressed representations for quantum many-body states and quantum circuit simulation, as strong classical baselines for quantum-advantage claims, and as quantum-inspired numerical tools in machine learning, optimization, finance, scientific computing, and data compression \cite{QAdvPerspective,MLTN2026}. In this sense, tensor networks are not only quantum-inspired algorithms but also part of the classical benchmark that any claim of practical quantum advantage must confront.

A further change is the maturation of the surrounding software stack. Contraction planning, GPU execution, and tensor-operation interfaces have become more standardized and more relevant for deployment, which makes implementation details increasingly important in industrial assessments \cite{TAPP}. Across all these roles, the central limitation remains structural: tensor-network methods are effective when the data, model, operator, or feasible set admit controlled ranks and manageable contraction widths. They are therefore not generic black-box cures for high dimensionality or NP-hardness.

\section{Introduction to tensor networks}
The first thing we will do is to make a brief introduction to tensor networks in order to understand the subsequent sections. In an effort to stay within the main objective of the paper, we will assume knowledge of basic linear algebra concepts and focus on tensor networks as such. Complementary information on tensor networks can be found in \cite{TN,TN2}.

\begin{figure}[htb] 
\begin{center} 
  \includegraphics[width=7.5cm]{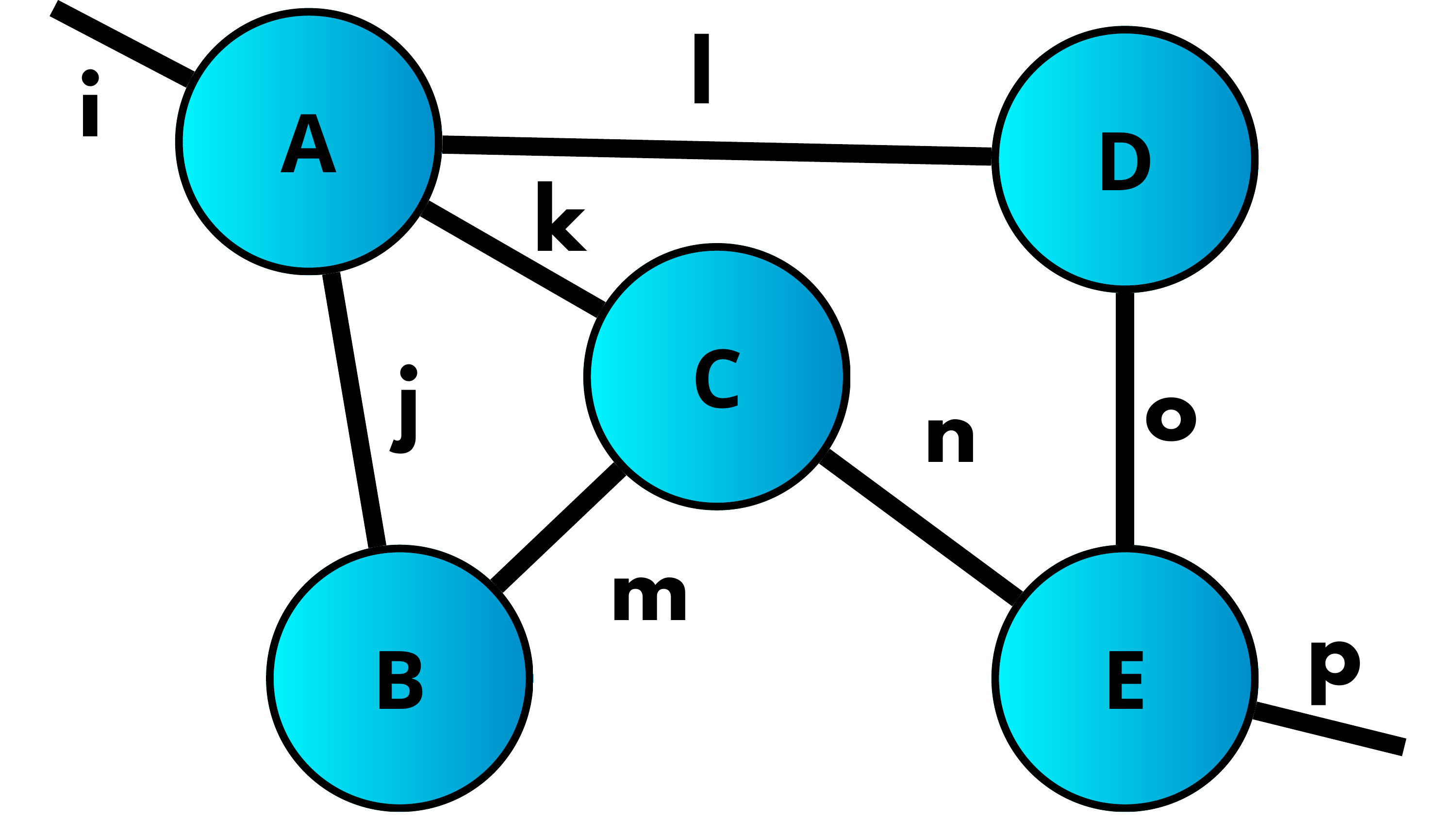}
\end{center} 
\caption{Tensor network representing the 2-tensor $T_{ip}$ of Eq. \ref{eq: TN General}.} 
\label{fig: general TN} 
\end{figure}

A tensor network (TN) is a graphical representation of a multilinear equation between different tensors. We can see an example in Fig. \ref{fig: general TN}, where we have the tensor network representing the 2-tensor $T_{ip}$ whose elements are obtained by means of the contraction operation
\begin{equation}\label{eq: TN General}
    T_{ip} = \sum_{j,k,l,m,n,o} A_{ijkl}B_{jm}C_{kmn}D_{lo}E_{nop}.
\end{equation}

As can be seen, this is a very simple representation of a very complex operation, in whose larger generalizations it is complicated to determine the relationships between the tensors contracted and the most efficient way to contract them. In addition, we can further abstract the tensor network by eliminating the names of the indexes and tensors if we know by some rule what they should be. An example would be that the tensor indexes are placed according to the clockwise orientation.

The rules followed by such representations are as follows:
\begin{enumerate}
 \item Each node of the network represents a tensor.
 \item Each outgoing line of a tensor represents an index of that tensor. Therefore, a node with $n$ outgoing lines represents an $n$-tensor.
 \item Each line connecting two tensors represents that both tensors are contracted with respect to a common dummy index.
 \item Lines that only come out of one tensor, but do not connect to another, represent free indexes. Therefore, a tensor network with $n$-free indexes represents an $n$-tensor.
 \item Nodes that are not connected by an index are in tensor product with each other.
\end{enumerate}

\begin{figure}[htb] 
\begin{center} 
  \includegraphics[width=7.5cm]{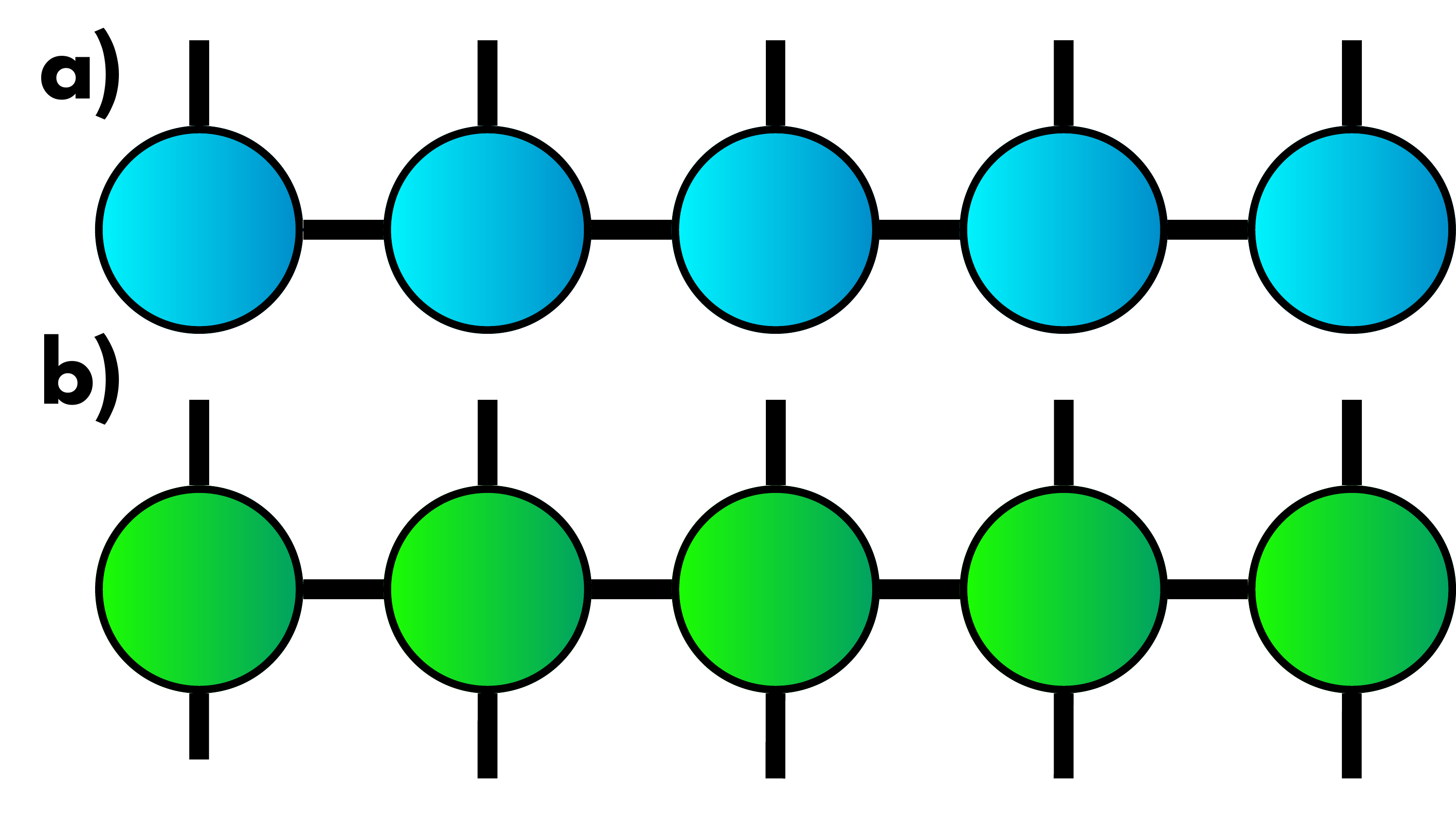}
\end{center} 
\caption{a) MPS form of a 5-tensor, b) MPO form of a 10-tensor.} 
\label{fig: MPS and MPO} 
\end{figure}

A highly interesting and useful type of tensor network is the already mentioned MPS \cite{MPS1}, since it consists of a one-dimensional decomposed representation of a tensor by means of the product of matrices. We can see an example in Fig. \ref{fig: MPS and MPO} a), where we see the MPS representation of a 5-tensor. This MPS representation is composed of 2 types of indexes: the bond indexes, which will have dimension $b$, and the physical indexes, of dimension $d$, corresponding to the dimensions of the tensor to be represented. The physical indexes are the free indexes, which coincide with the indexes of the tensor to be represented, while the bond indexes are those that connect the tensors of the network to each other and are responsible for accounting for the correlations between the physical indexes. It should be noted that there are other versions of the MPS representation with different boundary conditions (the way to link the end nodes), but the simplest is this one. There is also the possibility that the dimensions of the bond indexes are different, as are those of the physical indexes, for reasons we will mention at the end of this section.

For an $N$-tensor MPS representation, with constant bond dimension $b$ and constant physical dimension $d$, the number of elements to store is $2db + (N-2)db^2$ versus the $d^N$ that would be needed in the dense, contracted version of the tensor, $db$ for each end node and $db^2$ for each node in the interior of the chain.

Any tensor admits an exact open-boundary MPS/TT representation, although not always efficiently \cite{MPS1,TT_rounding}. The relevant question is how the bond dimensions scale with the tensor size and with the ordering of the indexes.

\begin{proposition}
Let $X\in\mathbb{F}^{d_1\times\cdots\times d_N}$, and let $X^{\langle k\rangle}\in\mathbb{F}^{(d_1\cdots d_k)\times(d_{k+1}\cdots d_N)}$ be the unfolding across the bipartition $(1,\dots,k)\mid(k+1,\dots,N)$. Then there exists an exact open-boundary TT/MPS representation whose minimal bond dimensions satisfy
\begin{equation}
    r_k=\operatorname{rank}\left(X^{\langle k\rangle}\right),\qquad k=1,\dots,N-1.
\end{equation}
Consequently,
\begin{equation}
    r_k\leq \min\left(d_1\cdots d_k,\ d_{k+1}\cdots d_N\right),
\end{equation}
and if $d_j=d$ for all $j$, then
\begin{equation}
    \max_k r_k \leq d^{\lfloor N/2\rfloor}.
\end{equation}
Generic tensors attain these maximal unfolding ranks.
\end{proposition}

\begin{proof}
Successive SVDs across the cuts $(1,\dots,k)\mid(k+1,\dots,N)$ produce an exact TT/MPS with bond dimensions $\operatorname{rank}(X^{\langle k\rangle})$. Conversely, cutting any exact TT/MPS between nodes $k$ and $k+1$ factorizes $X^{\langle k\rangle}$ through the bond space of dimension $r_k$, so $\operatorname{rank}(X^{\langle k\rangle})\leq r_k$. Hence the minimal bond dimension on that cut is exactly $\operatorname{rank}(X^{\langle k\rangle})$.
\end{proof}

Therefore, exact MPS representations can still be exponentially large, since for uniform physical dimension $d$ one may need $b\geq d^{\lfloor N/2\rfloor}$. Efficient tensor-network methods rely instead on low effective ranks or on controlled truncation. For example, in academic contexts a widespread use is to represent quantum states of $N$ particles, that is, state vectors of $d^N$ components.

An extension of the MPS representation is the Matrix Product Operator (MPO) \cite{MPO}, consisting in allowing each node to have 2 physical indexes, as we see in Fig. \ref{fig: MPS and MPO} b). The usefulness of this representation lies in cases where we want to represent a matrix of $d^{2N}$ components, such as a quantum interaction. In this case, the number of elements to store is $2d^2b+(N-2)d^2b^2$.

Such forms can be obtained mainly in three ways. The first is by logical construction of the representation, so that the interactions between the different nodes of the tensor network generate the tensor we want. The second will be the variational construction, consisting of starting with a test set of tensors and using techniques such as gradient descent to get it to become the one we want to represent. The third is by iterative singular value decomposition (SVD) \cite{SVD,TT_rounding}, where we reshape the tensor across successive bipartitions and apply SVDs recursively. There is also an optimized method for sparse tensors \cite{TT_Sparse}.

Importantly, either representation can be used to represent a vector, matrix, or general tensor, since we can convert an $m$-tensor to an $n$-tensor by techniques such as `grouping' and `splitting' \cite{TN}, a bijective mapping from one set of indexes to another set of indexes.

Consider a matrix $A\in\mathbb{F}^{P\times Q}$ that we want to represent as an MPO with $n$ nodes. To do this we first split the row and column indexes according to factorizations
\begin{equation}
    P = \prod_{k=1}^{n} p_k,\qquad Q = \prod_{k=1}^{n} q_k.
\end{equation}
This identifies the global row and column indexes with tuples
\begin{equation}
    a\leftrightarrow (a_1,\dots,a_n),\qquad b\leftrightarrow (b_1,\dots,b_n),
\end{equation}
where $a_k\in \{1,\dots,p_k\}$ and $b_k\in \{1,\dots,q_k\}$, and therefore
\begin{equation}
    A_{a,b}=\mathcal{A}_{a_1,\dots,a_n,b_1,\dots,b_n}.
\end{equation}

\begin{figure}[htb] 
\begin{center} 
  \includegraphics[width=7.5cm]{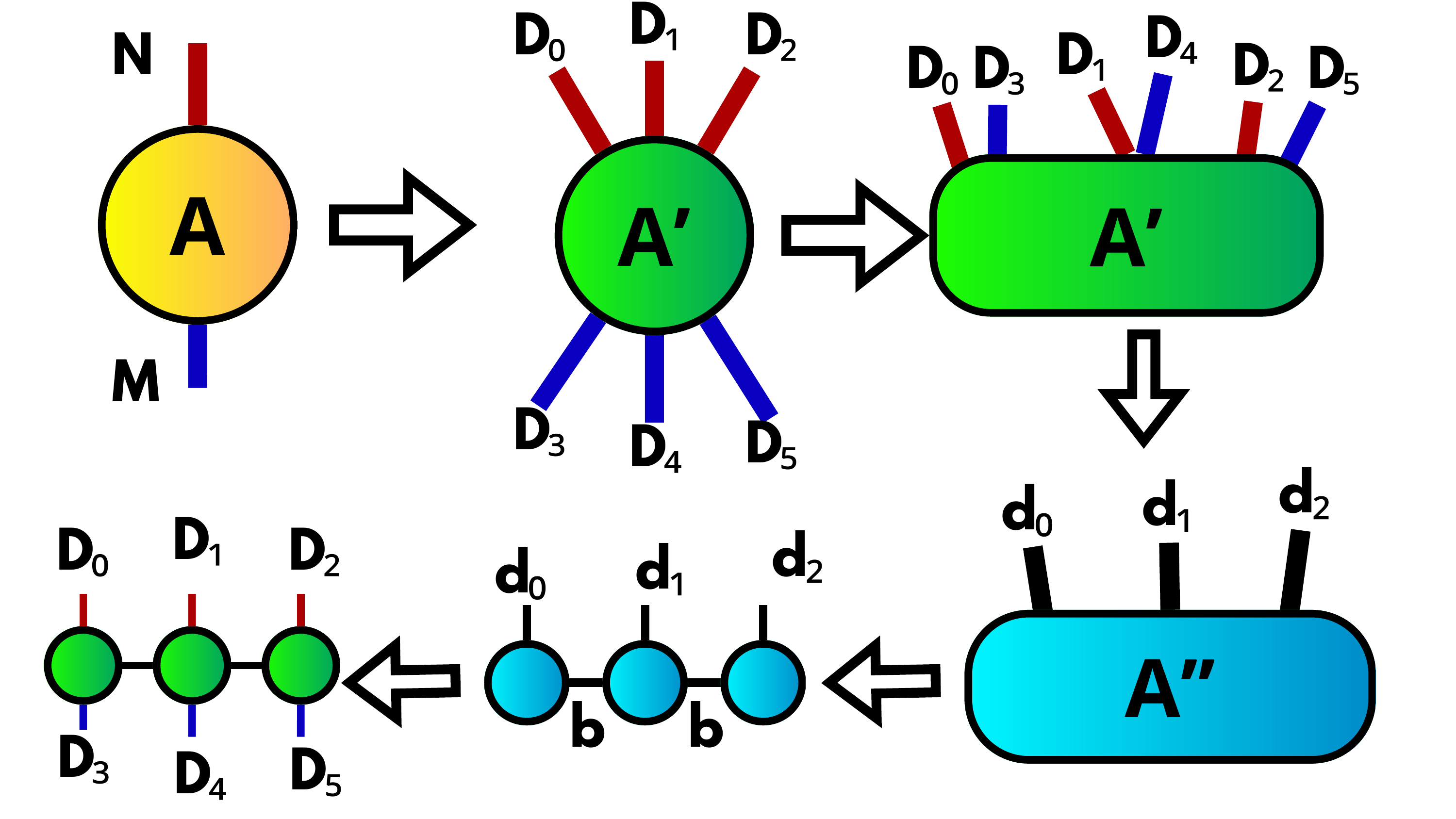}
\end{center} 
\caption{Process of compression of a matrix to its MPO representation. 1) We apply a splitting. 2) We join the indexes of each node in pairs. 3) We apply a grouping of the index pairs. 4) We perform the iterative SVD. 5) We apply a splitting to each physical index.} 
\label{fig: Compression_Layer} 
\end{figure}

To construct the MPO, we permute the indexes to $\widetilde{\mathcal{A}}_{a_1,b_1,a_2,b_2,\dots,a_n,b_n}$ and group each pair $(a_k,b_k)$ into a local index of dimension $d_k=p_kq_k$. We then apply TT-SVD to this reordered tensor and finally split each local index again to recover the two physical indexes per MPO node. In the notation of a $2n$-tensor with dimensions $(D_0,\dots,D_{2n-1})$, if the first $n$ dimensions correspond to the rows and the last $n$ to the columns, then the correct local dimensions are
\begin{equation}
    d_i=D_iD_{n+i},\qquad i=0,\dots,n-1.
\end{equation}
The formula $d_i=D_{2i}D_{2i+1}$ is valid only after explicitly interleaving the dimensions as $(D_0,D_n,D_1,D_{n+1},\dots,D_{n-1},D_{2n-1})$. The whole process can be seen in Fig. \ref{fig: Compression_Layer}.

If instead of wanting an exact representation of a tensor, we would like an approximate representation of it in order to have a smaller $b$ and thus save memory, we can take advantage of these two representations. In the variational construction we will simply operate with smaller $b$, while in the iterative SVD we will perform truncated SVD (TSVD) \cite{TSVD}, keeping only the largest singular values on each unfolding. We can also approximate a previously obtained MPS representation \cite{TT_rounding}. In TT-SVD, if the discarded squared singular-value mass at step $k$ is $\epsilon_k^2$, then the resulting approximation $\widetilde{X}$ satisfies
\begin{equation}
    \|X-\widetilde X\|_F^2 \leq \sum_{k=1}^{N-1} \epsilon_k^2 .
\end{equation}
Thus, choosing $\epsilon_k=\epsilon/\sqrt{N-1}$ yields $\|X-\widetilde{X}\|_F\leq \epsilon$, and for fixed TT ranks the construction is quasi-optimal up to a factor $\sqrt{N-1}$ in Frobenius norm. The accuracy of this approximation will depend both on the tensor itself that we want to represent and on the mapping we use, since if we put together in the same nodes the most correlated `columns', all their correlations will be within the node itself. Moreover, bringing strongly correlated indexes close to each other can reduce the bond dimensions needed for a good approximation. For injective homogeneous MPS, correlations of local observables decay exponentially with distance when the transfer matrix has a nonzero spectral gap. More generally, the MPS topology constrains entanglement across cuts through the bond dimensions, which is why the ordering of variables affects the ranks needed for an accurate representation.

\section{Basic tensor properties and more complex techniques}
In this section we will make a connection between the basic properties of tensor networks and certain more advanced and complex techniques. In this way we will create a bridge between the abstraction of tensor networks and their practical applications. To do so, we will approach each case explaining which properties are determinant in it and which is the advantage it offers with respect to other methods.

\subsection{Simulation of quantum systems}
As is evident from their historical context, tensor networks are especially intended for the efficient representation of many-body physical systems. This is of great interest in fields such as condensed matter physics \cite{Many1,Many2} or quantum computing \cite{Noisy_QC,TN_QC_Circuit}.

For this purpose an estimated (or `ansatz') solution that is representable as a tensor network is chosen and worked with, translating into tensor notation also physical operators and magnitudes. This allows us to work with a limited amount of memory, unlike the case of working with the entire statevector. Cases where tensor networks are very successful are in obtaining minimum energy states in quantum chemistry \cite{Q_Chemistry} and in the time evolution of certain quantum states to obtain certain magnitudes and properties of them \cite{Q_Adv1,Q_Adv2}.

Beyond many-body simulation, tensor networks have become a central tool for assessing claims of quantum advantage. Several recent quantum-hardware experiments have been analyzed by comparing them with optimized tensor-network simulations \cite{Q_Adv1,Q_Adv2,QAdvPerspective}. This does not remove the relevance of quantum hardware; rather, it raises the standard for advantage claims and clarifies the regimes in which classical contraction methods remain competitive.

\subsection{Machine learning model compression}\label{ssec: mach}
The best known application of tensor networks is the compression of machine learning models \cite{Compress1,Compress2,ML_TN}. This can be done either by compressing an already created model and training it or by creating a compressed model and training the tensor network that represents it directly. The key to model compression lies in finding an efficient representation of the model that preserves its analysis properties. We will explain how it works by compressing a dense layer in a neural network to its approximate MPS and MPO representations.

To begin with, we will have to keep in mind that the action of a dense layer of a neural network, prior to the action of an activation function, can be written as

\begin{equation}
    \vec{v} = A\vec{x} +\vec{c},
\end{equation}
where $A$ is a matrix of dimension $N\times M$, $\vec{v}$ and $\vec{c}$ are vectors of dimension $N$ and $\vec{x}$ is a vector of dimension $M$. The input to the layer is the vector $\vec{x}$, while the output will be the vector $\vec{v}$. The layer has $NM$ trainable parameters in the matrix $A$ and $N$ trainable parameters in the bias vector $\vec{c}$, for a total of $NM+N=N(M+1)$ parameters before compression. To reduce the number of parameters we can represent the matrix $A$ as an approximate MPO form with a certain bond dimension $b$ so that we have a smaller number of parameters than in the original dense version. We can do the same with the vector $\vec{c}$ obtaining its approximate MPS form. In this way, we have compressed the model layer. This is favorable both to save memory space and to train the model faster and avoid problems such as overfitting, depending on the representation used and the original model.

Since the $A$ matrix approximation and the $\vec{c}$ vector approximation can be made arbitrarily close to the originals by increasing the $b$ bond dimensions, we can always start from the original model and compress it until we saturate a preset tolerable error. However, we can also compress the elements of the model and run it to see how accurate it is with respect to the original model, increasing the compression until we reach that maximum error.

Another possibility would be to use a compressed version of the untrained model, with a controlled initialization as in the original, and train the compressed model directly. This has the advantage that we do not lose accuracy by approximating our model, but train an already pre-approximated model. We can also initialize the model by compressing an already trained model and retraining it.

There are many compression schemes in addition to the MPS and MPO, such as the PEPS \cite{PEPS,PEPS2} or the MERA \cite{MERA}. These allow certain operations to be performed efficiently on their compressed representations and can be optimized with local variational methods, including DMRG in the MPS/one-dimensional setting and related sweep or variational algorithms for PEPS, MERA and other tensor-network geometries \cite{MERA,DMRG}.

From 2024 onward, much of the applied focus has shifted from compressing relatively small neural networks to compressing transformer and large language model components. MPO/TT decompositions can be applied to dense or linear layers, attention projections, and feed-forward blocks, while tensorized adapters provide parameter-efficient fine-tuning mechanisms \cite{Compress2,MLTN2026,MetaTT,PicoGPTMPO}.

However, parameter compression alone does not guarantee lower wall-clock latency. Practical acceleration requires training and inference to operate directly on the compressed cores, together with efficient contraction kernels; reconstructing dense matrices removes much of the benefit. For industrial AI, the relevant question is therefore not only the compression ratio, but the full deployment path: whether the compressed cores are used natively during training and inference, whether hardware kernels support the resulting contractions, and whether the accuracy--latency--memory trade-off improves over quantization, pruning, distillation, LoRA-type adapters, and conventional low-rank decompositions \cite{Compress2,MetaTT,PicoGPTMPO,TensorKrowch}.

\subsection{Application of high dimensionality operations}\label{ssec: high dim}
A problem that we may have in real industrial cases is the need to apply high dimensional operations, such as creating certain kernels for methods like the Support Vector Machine \cite{SVM}. We can take as an example wanting to apply a kernel in tensor product of the components of an $\vec{x}$ input vector of $N$ components \cite{Anomaly}. This kernel is
\begin{equation}\label{eq: product kernel}
    \phi(\vec{x})_{i_1, i_2, \dots, i_{N}} = \phi_1(x_1)_{i_1} \phi_2(x_2)_{i_2} \dots \phi_{N}(x_{N})_{i_{N}},
\end{equation}
where each local feature map $\phi_j(\cdot)$ acts on the $j$-th component of the vector $\vec{x}$. In this case, we can consider
\begin{equation}
    \phi_j(x_j)_0=x_j,\qquad \phi_j(x_j)_1=1.
\end{equation}
Then every component of the global feature map is
\begin{equation}
    \phi(\vec{x})_{i_1,\dots,i_N}=\prod_{j:\,i_j=0}x_j,
\end{equation}
so the resulting monomials are square-free: each variable appears at most once. The induced kernel between two inputs is
\begin{equation}
    K(\vec{x},\vec{z})=\langle \phi(\vec{x}),\phi(\vec{z})\rangle=\prod_{j=1}^{N}(1+x_jz_j).
\end{equation}
Therefore the global feature map gives all possible square-free products between the components of the input vector, as a $2^N$-component tensor that expresses correlations in the vector. As we can see, in the standard way this would require an exponential amount of memory and time in the size of the input vector. Furthermore, if we want to apply a matrix to such a feature vector we would need a matrix $A$ of dimension $M\times2^{N}$, which scales prohibitively.

\begin{figure}[htb] 
\begin{center} 
  \includegraphics[width=7.5cm]{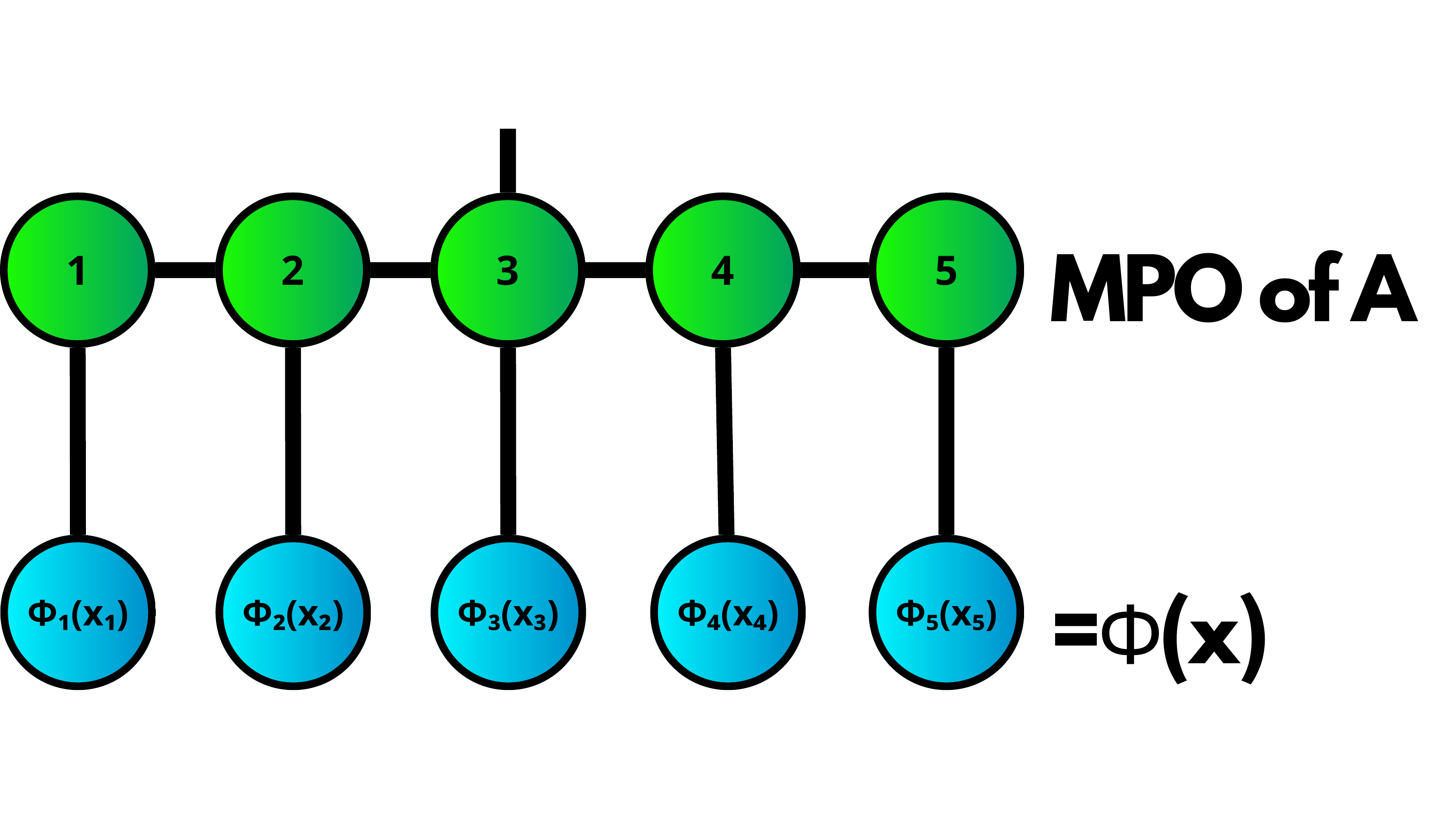}
\end{center} 
\caption{Eq. \ref{eq: product kernel} and MPO layer for an input vector of $N=5$ components.} 
\label{fig: Kernel_and_MPO} 
\end{figure}

However, by taking advantage of the properties of tensor networks we can use an MPO representation for such a matrix, of the form depicted in Fig. \ref{fig: Kernel_and_MPO}, provided that the output dimension $M$ is either small enough to be treated as a single physical output index or factorized as $M=\prod_k p_k$, so that $A$ can be represented with local input dimension $2$ and local output dimensions $p_k$. The advantage is lost if the required MPO bond dimensions grow exponentially. Under these conditions the operation can be performed through a sequence of local contractions. If the MPO has local input dimension $2$, local output dimensions $p_k$, and bond dimensions $\chi_k$, contracting it with the product feature map produces a compressed output with cores of size roughly $\chi_{k-1}\times p_k\times \chi_k$, at a cost polynomial in the local dimensions and bond dimensions. If the final $M$-dimensional output vector is materialized explicitly, an additional cost at least linear in $M$ is unavoidable. Thus the advantage comes from keeping the intermediate and, when possible, the output representation compressed.

\begin{figure}[htb] 
\begin{center} 
  \includegraphics[width=7.5cm]{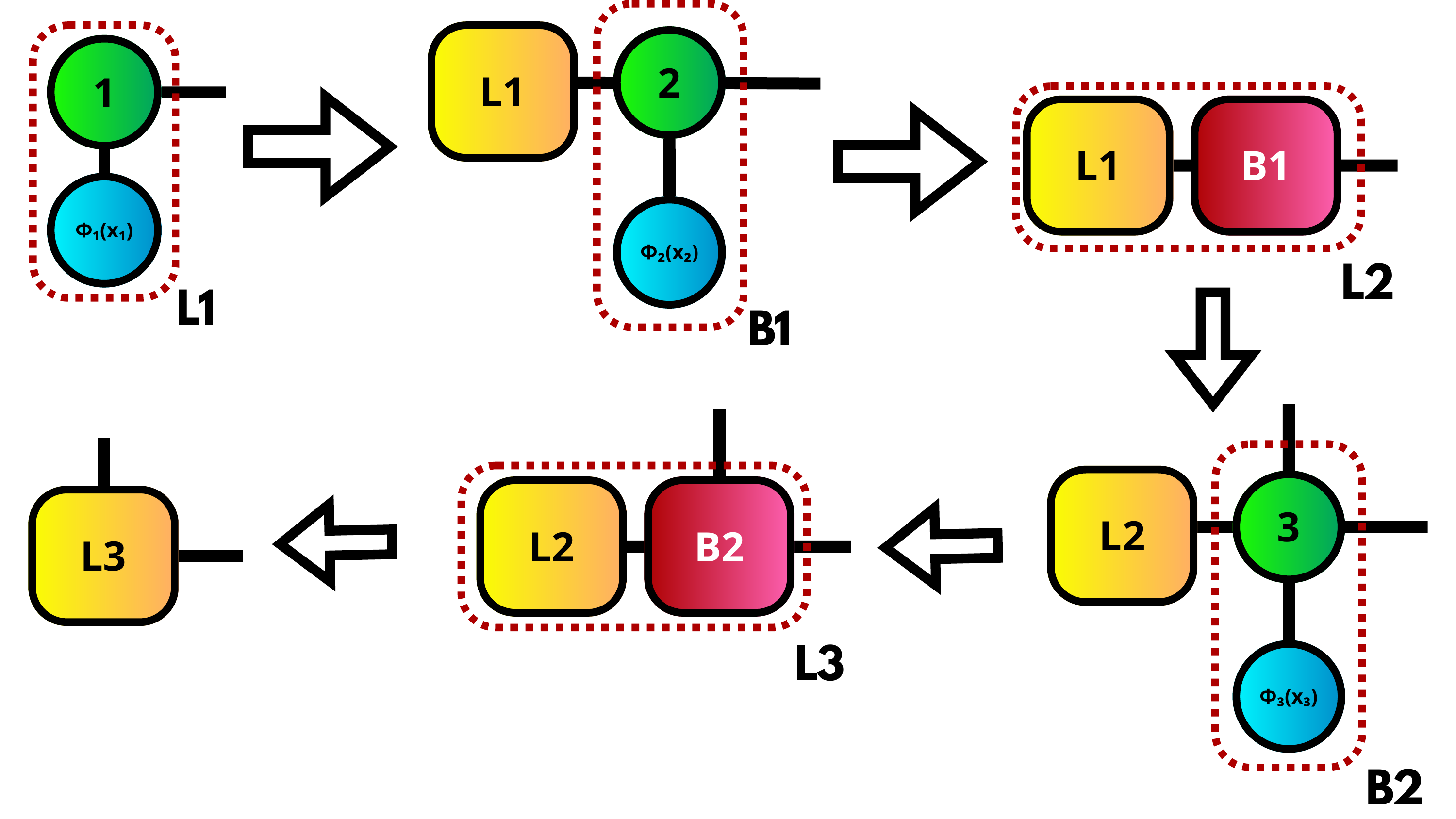}
\end{center} 
\caption{Contraction of the first three nodes of the MPO layer with the first three nodes of the kernel. 1) We contract a kernel node with its corresponding MPO layer node. 2) We contract the next kernel node with its layer node. 3) We contract the two new nodes obtained. 4) We repeat the contraction process for the next pair of nodes. 5) We contract the current nodes, maintaining the physical index.} 
\label{fig: Contraction} 
\end{figure} 

With this methodology we can apply an operation in which we weight elements of type $x_1x_2x_3x_4x_5$, $x_1x_2x_3x_4$, $x_1x_2x_3$, $x_1x_2$ and $x_1$, efficiently and with a reduced number of elements.

\subsection{Data compression and denoising}
In the field of big data there is a need to work with large volumes of data, which can be very expensive or unfeasible for certain systems. The tensor compression property that we have seen in representations such as MPS allow us to compress \cite{Big_Data} and handle \cite{Big_Data2} large amounts of data. To do this we must first tensorize the data and then compress it. The tensorization step is not neutral: different reshapes, quantized tensor-train encodings, or variable orderings can lead to very different ranks. Denoising and compression are effective when the signal has low effective tensor rank and the discarded components are dominated by noise; for dense global correlations or structured noise, the required ranks may remain high.

\subsection{Combinatorial optimization}
Solving combinatorial optimization problems typically requires runtimes that grow rapidly with problem size. Quantum and quantum-inspired methods can provide useful heuristic and structural approaches for certain instances, especially when the cost function or the constraints admit compact tensor-network representations. However, for general NP-hard problems no polynomial-time exact method is known, and tensor-network ranks may still grow exponentially in the worst case. There are different types of combinatorial optimization solvers.

The first one is the one developed in \cite{TTOpt}, running in a black box. This algorithm starts from a cost tensor, $T$, whose subscripts are the components of the solution vector. This tensor will have the cost of all possible states, although it is not necessary to know the tensor explicitly: it is enough to have a function that gives us the cost as a function of its indexes. The method builds a TT approximation by tensor cross interpolation guided by the \textit{max-volume} principle, that is, by selecting submatrices with large absolute determinant, or suitable rectangular generalizations, in order to obtain stable interpolatory approximations. This should not be interpreted simply as selecting singular values with the largest product. From this representation we will obtain, with high probability, the lowest cost element.

The problem with this algorithm is that when we work with problems with many invalid configurations, it tends to fail because it sees that almost all or all the elements are the same, since it is sampling the tensor. Therefore, we need an encoding, however complicated it may be, in which we reduce the space of invalid configurations, so that the algorithm sees a certain functional map. 

A recent direction is to build feasibility directly into the tensor network. Feasible tensor networks and constrained MPS encode constraints in the support or block structure of the state, so that samples or amplitudes correspond only to configurations satisfying the prescribed constraints \cite{FeasibleTN,ConsTraining}. This can reduce the burden of penalty tuning and avoids wasting probability mass on invalid configurations. The limitation is that constructing such networks may itself be problem-dependent, and the resulting ranks can still grow quickly for dense or global constraints.

The second possibility is based on the use of Generative Artificial Intelligence (AI) boosted with tensor networks \cite{GEO}.

The third possibility is the use of imaginary-time evolution on the quantum state in uniform superposition of all possible combinations, equivalent to the application of a Hadamard gate to each qubit of a quantum circuit. Let $\Omega$ be the set of feasible configurations, let $C(s)$ be the cost function, and let
\begin{equation}
    D_\beta=\exp(-\beta C),\qquad |\psi_0\rangle=\frac{1}{\sqrt{|\Omega|}}\sum_{s\in\Omega}|s\rangle.
\end{equation}
where $D_\beta$ is the diagonal operator defined by $D_\beta|s\rangle=\exp(-\beta C(s))|s\rangle$.
Then the normalized state
\begin{equation}
    |\psi_\beta\rangle=\frac{D_\beta|\psi_0\rangle}{\|D_\beta|\psi_0\rangle\|}
\end{equation}
has measurement probabilities proportional to $\exp(-2\beta C(s))$, so low-cost combinations are favored.

Imaginary-time evolution is not a deterministic closed-system unitary evolution. On quantum hardware it therefore requires additional mechanisms such as normalization, measurements, ancillas, postselection, or variational approximations. In classical tensor-network simulations, the non-unitary operator can be applied directly, provided that the resulting tensor-network ranks remain controlled \cite{Combin}. If $S_\star$ is the set of minimizers, $C_\star$ is the minimum cost, and $\Delta$ is the gap to the best non-optimal value, then
\begin{equation}
    \Pr(s\notin S_\star)\leq \frac{|\Omega|-|S_\star|}{|S_\star|}\exp(-2\beta\Delta).
\end{equation}
Indeed, the total unnormalized weight of non-optimal states is at most $(|\Omega|-|S_\star|)\exp(-2\beta(C_\star+\Delta))$, whereas the optimal states contribute at least $|S_\star|\exp(-2\beta C_\star)$, and dividing both bounds gives the claim. If we needed to take into account all possible combinations explicitly, we would need an unbearable amount of memory, resulting in a brute-force search. However, by combining the ability to compress information and simulate quantum states with the ability to apply non-unitary tensor-network operations, we can create such a state for various problems \cite{TN_TSP,TN_QUBO,TN_Task}. This can be done in an exact or approximate way, by recompressions, depending on how much memory we want to devote to it.

The main limitation is that for NP-hard problems we may still need an exponential amount of memory and time to solve them exactly, because the tensor-network ranks can grow exponentially in the worst case. By making intelligent adjustments in the encoding and expression of the problem, the creation and contraction of the tensor network, and by applying various simplifications in the constraints, we can nevertheless obtain useful approximate or heuristic algorithms on structured instances.

The fourth is the simulation of a physical system represented by a variational tensor network, with a cost function given as a Hamiltonian that we can write in MPS or MPO representation. In this case we will perform variational updates of the state tensor network, using techniques such as gradient descent, DMRG in the MPS setting, or related local sweep methods for other tensor-network geometries, in order to minimize the energy of the system and find the optimal combination. The effectiveness of these types of techniques depends very much on the problem to which they are applied and they can get stuck in local minima, especially for problems with constraints. In addition, there is the limitation that we must be able to express the cost function in a tensor network representation.

\subsection{Software, contraction planning, and deployment}
From an industrial perspective, the ansatz is only part of the engineering problem. Contraction-path optimization can be as important as the network architecture itself, because runtime and peak memory depend strongly on the chosen contraction tree, slicing strategy, and backend. Recent software ecosystems and standards, including ITensor/ITensors.jl workflows, quimb/cotengra-style contraction planning, TensorKrowch for machine-learning integration, GPU backends such as NVIDIA cuTensorNet, and community efforts toward standardized tensor-operation interfaces, make tensor-network prototyping and deployment considerably more practical than a few years ago \cite{TAPP,TensorKrowch,cuTensorNetDocs}. For reproducible industrial evaluation, it is therefore important to report the tensorization, bond dimensions or truncation tolerances, contraction-path optimizer, hardware backend, numerical precision, peak memory, and wall-clock time, together with the accuracy metric relevant to the application.

\section{Use case studies}
In this last section we present use cases in which tensor-network techniques may have direct industrial applicability.
\subsection{Finance}
In the contemporary financial domain, high dimensionality emerges as a challenging phenomenon. With rapid technological advancement and the availability of large-scale data, financial markets have become inherently multidimensional. Thanks to this, in-depth market analysis is possible.

\subsubsection{Portfolio optimization}
Portfolio optimization consists of determining the best allocation of financial assets in which to invest in order to maximize a cost function, which is usually one that maximizes return and minimizes risk. This problem is already complicated in a static way, becoming extremely complex in a dynamic way, with a portfolio to be optimized over a series of trading days. The high dimensionality and constraints of the problem make it generally an intractable problem in an exact manner.

Tensor networks can be used to explore the space of inversion solutions in order to search for the optimal combination. In \cite{Portfolio} we can see how they use imaginary time evolution on an MPS representation to solve the optimization problem. They achieve sharpe ratio results, a measure of excess return per unit of risk on an investment, superior to those obtained by other methods for large sizes, although with lower payoff and much longer run time for large problems than Dwave's hybrid solver.

\subsubsection{Interpretable predictions}
In \cite{Prediction}, a recurrent neural network method is presented for making interpretable predictions, those in which the underlying reasoning can be understood, using an MPS representation of the recurrent layer weight matrix. It can be observed that the accuracy, Sharpe and return results are higher in the tensor network version than in the other compared versions.

\subsubsection{Option pricing and risk simulations}
Tensor trains, MPS, and QTT representations are increasingly used for option pricing, Greeks, and large-scale risk revaluation. The motivation is that full-grid PDE or tree-based methods suffer from dimensionality, while Monte Carlo methods may require many samples for stable Greeks or tail-risk estimates. Tensor-network methods can compress price surfaces, payoff tensors, operators, or parameter dependence, often through TT-cross or QTT representations \cite{OptionPricingTN,MultiAssetOptionsTN,GreeksTT,TTOptionParam,STNGPROption}.

The practical caveat is that the gains depend strongly on smoothness, parameter dimension, rank growth, calibration cost, and on whether the underlying pricing engine can be queried efficiently. In industrial settings, these methods should be compared against strong Monte Carlo, sparse-grid, Fourier, and reduced-order baselines for the target asset class and accuracy regime.

\subsection{Medicine}
The field of medicine is one of the fields with the highest dimensionality, both for the diagnosis and treatment of diseases, as well as in the design and production of drugs. Because of this, there has recently been a great interest in the application of Artificial Intelligence techniques in this field, obtaining great results.

\subsubsection{Drug discovery}
Drug simulation is a monumental challenge in contemporary science due to the complexity of molecular interactions and the enormous number of possible configurations. Fault-tolerant quantum computing may eventually provide advantages for selected electronic-structure and quantum-chemistry tasks, but in the near term tensor-network methods such as DMRG and related ansatz-based techniques remain classical tools for exploiting structure in correlated molecular systems. Their usefulness depends on orbital ordering, entanglement structure, active-space selection, and the ranks required for the desired accuracy.

Similarly, we can save time and costs in drug discovery by predicting new therapeutic properties associated with new drugs. This is a data-driven tensor-decomposition and tensor-completion use case rather than a quantum many-body simulation \cite{Drug}. To this end, we create a drug-gene-disease tensor, which accounts for the relationships found between these drug-gene, drug-disease and gene-disease pairs. With this, we formulate a tensor completion problem. The initial tensor will have only the direct relationships of the pairs we have found, while the rest of the more complex relationships will be the ones we have to determine. For this, we will perform a generalized tensor decomposition (GTD), to model part of these relationships, and we will apply a multilayer perceptron (MLP) to capture the rest of the relationships.

The main limitation of this method is that its predictive capacity is inferior for new diseases, so other techniques must be coupled to it.

\subsubsection{Medical image analysis}
The analysis of images, and especially medical images, is a very complicated task, either because of the characteristics of these images, the noise in them or the high dimensionality that the problem can reach. Tensor networks can be useful in high-dimensional and noisy imaging tasks when the chosen tensorization exposes low-rank, local, or multiscale structure, and when truncation removes noise without discarding clinically relevant signal.

In \cite{Imagemed} authors present methods for classifying medical images, using exactly the scheme we explained in Section \ref{ssec: high dim}, interpreting the images as an input vector, a trigonometric feature map, and the MPO layer as a classifier. Thus, by contracting the tensor network we will obtain a vector that will tell us the probabilities of that image belonging to one group or another. This is done in combination with other more advanced machine learning techniques and finally we obtain a high level of compression with levels of area under the ROC (Receiver Operating Characteristic) curve similar or superior to the methods with which it is compared. All this with much lower GPU requirements.

Medical image segmentation is vital for disease diagnosis, treatment monitoring, surgical guidance, experimental drug development and medical education. A method for medical image segmentation with tensorized transforms, introduced in the paper, is shown in \cite{Image3d}. In this method, a tensorized version of the self-attention module, a step of the transformer, is realized.

\subsection{Simulation of quantum materials and topological materials}
Simulation of quantum materials and topological materials is an extremely complex problem, because they are high-dimensional problems. Quantum many-body systems require an exponential memory in the number of elements to represent their wave functions. The utility of these materials is enormous, given their curious applications, such as high-speed electronics, solar cells, quantum sensors, telecommunications and quantum computing.

To deal with this, concrete ansatz are proposed
\cite{Materials,Topology1,Topology2,PEPS_Methods} which accurately and efficiently represent the minimum energy states of these systems. We start with an initial state and perform various transformations, such as DMRG in one-dimensional/MPS settings or related variational and imaginary-time algorithms for other tensor-network geometries, to bring it down to its minimum energy state. Once this state is obtained, we can easily obtain properties of the state by applying the tensors of the quantities to be measured.

\subsection{Battery modelling and system identification}
Battery modelling is an industrially relevant application where tensor methods can appear not only through quantum materials simulation, but also through data-driven system identification. The works considered here use tensor-network-based Volterra models for lithium-ion batteries, including MIMO settings, to represent nonlinear dynamical input-output relationships in compressed form \cite{Batteries,Batteries2}. In this setting, the tensor network is not an ansatz for a many-body ground state; it is a structured representation of a high-order nonlinear model. The main industrial value is compact modelling of battery dynamics for estimation, control, and diagnostics, while the limitations are the availability and representativeness of data, the stability of the identified model, and the rank growth required to capture nonlinear behaviour.

\subsection{Optimization}
Optimization problems in industrial environments are very important, as they may require the optimization of the order in which processes are performed, such as the assignment of times and tasks to machines in the Job Shop Scheduling Problem (JSSP). However, due to the problem size in these cases, it is very expensive to solve such instances due to the high number of combinations and the complexity of the cost functions.

\subsubsection{Route Optimization}
The Traveling Salesman Problem (TSP) is a widely studied historical problem, in which a route must be chosen that traverses all nodes of a certain network only once and offers the lowest cost associated with it, given by the costs of going from one node to another. Generalizations and particular cases of this problem model sharing problems, which are very important in industry.

For this purpose, \cite{TN_TSP} propose a TN method that can solve generalized TSP cases, with different variants. This method takes as variable the node in which one is at each step of the route and thus model the cost function as a Quadratic Unconstrained Discrete Optimization (QUDO) to the nearest neighbor with non-repetition constraints. Its basic mechanics resides in the imaginary time evolution described in \cite{TN_QUBO} and the connection of a set of layers that impose the constraints, eliminating non-compatible combinations in the represented state. By modifying such constraint layers and the QUDO modeling, different generalizations can be addressed, such as being able to repeat up to $N$ times a certain node, that the start and end can be different or that we can only pass through each group of nodes once. All these cases are studied in \cite{TN_TSP}.

A complementary 2026 direction is to use tensor-network generative models for TSP. In TN-GEO, an MPS Born machine is trained over valid tours using permutation-based variables and autoregressive masking, so that generated samples are feasible by construction rather than obtained through an $N^2$ binary encoding with penalty terms \cite{TNGEO_TSP}. This is a heuristic sampling strategy rather than a worst-case exact solver, but it illustrates an industrially relevant shift from penalty-heavy encodings to feasibility-aware sampling.

The main limitation of these methods is the exponential scaling induced by the constraints in the general case. Iterative or hybrid procedures such as the one explored in \cite{TN_Task} can mitigate this scaling on the tested structured instances, but no general worst-case polynomial guarantee is implied.

\subsubsection{Post assignment optimization}
A particular case of generalized TSP is the assignment of jobs to workers. Assume a set of workers who are already assigned to a set of jobs and we have a set of vacancies. The objective will be to redistribute workers to the set of jobs, including filled and vacant jobs, such that we maximize a certain cost function and there are only 0 or 1 workers per job. This problem has been studied and treated at the end of \cite{TN_TSP}, using the formulation of \cite{JRP}, which seeks to maximize a combination between the productivity of the job and the compatibility of the worker assigned to it. As in the route-optimization case, the practical performance depends on the structure of the instance and on the ranks required by the encoding.

\subsubsection{Manufacturing Sequence Optimization}
The optimization of manufacturing sequences is a very useful industrial problem. We have a set of products to be manufactured sequentially, such that there is additional manufacturing time when moving from manufacturing one type of product to manufacturing another. This time can be a time to change parts of a machine, configuration or materials. Therefore, we want to obtain the sequence that minimizes this overall additional time.

This case is studied in \cite{TN_TSP}, since it can be seen as a TSP in which we can be at each node as many times as there are outstanding products of the class associated to that node in the initial set. With this perspective, a node represents a class rather than a product to optimize the constraint layers.

\subsubsection{Machine task assignment optimization}
Optimal assignment of tasks to machines can be modeled as the choice of task execution on a set of machines such that there is a constraint between the tasks performed by one and another. An example of a constraint would be the following: `if machine 1 does a cutting, machine 2 cannot do a peeling'. These constraints can be extracted from a historical record or from the business knowledge of the usual machine manager. In addition, each task will have its own execution time on its corresponding machine.

This problem is addressed in \cite{TN_Task}, where a TN method with imaginary time evolution and constraint layers like those of \cite{TN_TSP} is combined with a genetic algorithm and an iterative method to reduce the time and memory scaling of such algorithm, taking advantage of properties of this type of problems. The main limitation of the method is that, although it has a high success rate on the tested structured instances, for certain cases it cannot obtain satisfactory solutions for a given memory limit, and no general worst-case polynomial guarantee follows.

\subsection{Big Data}
As its name indicates, in the field of big data large volumes of data are handled, so we need specific systems to deal with them.

The work \cite{Big_Data} presents different methods to deal with big data problems using tensor network decompositions, as well as other topics. On the other hand, in \cite{Big_Data2} a tensor network method for secret sharing in big data is presented, based on the decomposition into an MPS or a Tucker decomposition \cite{Tucker} of the original data tensor by means of consecutive TSVDs. Between each TSVD a tensor perturbation will be performed, by multiplying the new node obtained from the SVD by a $\Delta$ matrix, and the tensor to be decomposed in the next step by $\Delta^{-1}$. This achieves that the global tensor is the same, but the nodes of the representation are different. Finally, the first node of the representation is randomized. In this way, parts of the dataset can be stored in a distributed way while restricting reconstruction of the original data under the assumptions of the specific secret-sharing protocol. A claim of perfect secrecy would require an explicit adversarial model and a proof, for example in terms of statistical independence or zero mutual information between unauthorized shares and the secret. At the same time, it presents a way to perform distributed computing for big data.

\subsection{Classification}
As we have seen in the case of medical image classification, using methods such as the one presented, different types of images or general data can be classified.
In addition, there are more general methods. On the one hand, the inspiration of quantum circuits in tensor networks methods to make classification circuits \cite{ImageTN}. In this case, hierarchical trees are addressed.

On the other hand, we have the use of PEPS for image classification, where results similar to those of GoogLeNet, VGG-16 and AlexNet \cite{ImageTN2} are achieved by means of simple coupled tensor network systems.

\subsection{Artificial Intelligence}
As discussed in Section \ref{ssec: mach}, tensor networks can compress or parameterize components of artificial-intelligence models, especially when the chosen factorization aligns with the structure of linear layers or attention blocks. For industry, this is relevant for edge or on-premise deployment, privacy-sensitive environments, and reduced memory footprints, not only for parameter counting \cite{Compress2,MLTN2026,MetaTT,PicoGPTMPO}.

Any practical evaluation should therefore compare tensor-network compression or tensorized adapters against quantization, pruning, distillation, LoRA/PEFT, and conventional low-rank adapters, reporting accuracy, compression ratio, latency, throughput, energy use, and memory rather than parameter count alone.

We can also realize privacy-preserving machine learning techniques using tensor networks \cite{Privacy}. This is of vital importance when working with sensitive data, such as medical records, but the same deployment-level benchmarking requirements apply.

\subsection{Scientific computing, PDEs, and fluid dynamics}
Scientific computing is an industrially important area in which tensor networks are increasingly used as compressed representations of operators, solution fields, flow maps, and parameter-dependent objects. Recent examples include MPS/MPO approximations of PDE flow maps, tensor-network fractional-step solvers for incompressible flow, and TT representations of high-dimensional combustion manifolds for CFD \cite{PDEFlowMaps,TTFlowCurvilinear,FlameletTT}.

The attraction is clear: high-dimensional operators and lookup manifolds can sometimes be manipulated in compressed form rather than on full grids. The practical limitation is equally clear: performance depends on smoothness, separability, variable ordering, low rank, boundary and geometry handling, numerical stability, and comparison against strong classical reduced-order and sparse-grid baselines.

\subsection{Cybersecurity}
Cybersecurity is a very complex field, in which extraordinary measures are needed to protect users' privacy and detect possible attacks \cite{CiberTN}. Here we can encompass the techniques already seen above applied to this context.

\subsection{Anomaly detection}
Anomaly detection is a problem consisting of identifying unusual cases in a data set. This problem is very useful in detecting computer attacks. Because anomalous data can present many structures, having a large space of possibilities, it is required to model in a very effective way the normal data so that they can be identified with respect to the anomalous ones. This usually requires analyzing high dimensional data or applying very complex operations to it.

The work \cite{Anomaly} discusses a supervised method precisely to be able to deal with this problem. By applying an ansatz product like the one we explained in Section \ref{ssec: high dim}, and an operation in variational MPO that we will train, we can project the normal data to a hypersphere. In this way, the normal points will remain on the surface of the sphere, while the anomalous ones will go to the center or to the outside of the sphere. Moreover, the process can be performed very efficiently due to the structure of the tensor network to be contracted, scaling remarkably well with the size of the problem.

It has been tested with different datasets against different algorithms, giving good results. Its main limitation is that it is a supervised algorithm, in which we require a well-performed labeling prior for it to work properly. Even so, it can be extended to be part of an unsupervised algorithm by adding automatic pre-labeling.

\section{Conclusions}
We have seen that tensor networks are a versatile set of techniques for industrial contexts, with a growing literature and several promising application areas. We have explained the main features that make them attractive for structured high-dimensional problems and machine learning. We have also highlighted several applied cases and some of the techniques that have been used in the state of the art to address them, together with their advantages and limitations.

From an industrial perspective, tensor-network methods are most promising when the data, cost function or model exhibits locality, hierarchical structure, low effective entanglement, or correlations that can be aligned with the chosen network topology. They are less suitable when correlations are dense and global, when constraints induce high-rank projectors, or when exact guarantees are required for worst-case NP-hard families.

As of 2026, the industrial relevance of tensor networks is strongest in structured high-dimensional settings: quantum simulation and benchmarking, compressed machine-learning models, constrained optimization with exploitable structure, financial pricing and risk surfaces, scientific computing, and data-driven system identification. The main practical bottlenecks are tensorization choices, rank growth, contraction width, optimization stability, and hardware and software support. Consequently, industrial adoption should proceed through benchmarked pilots against strong classical baselines rather than through generic claims of exponential compression.

%Bibliography
\bibliographystyle{unsrt}  
\bibliography{references}

\end{document}